# Proper curvature collineations in non-static plane symmetric space-times

Ghulam Shabbir and M. Ramzan

Faculty of Engineering Sciences,

GIK Institute of Engineering Sciences and Technology,

Topi, Swabi, NWFP, Pakistan.

Email: shabbir@giki.edu.pk

**Abstract**

The most general form of non-static plane symmetric space-times is considered to study proper curvature collineations by using the rank of the $6\times 6$ Riemann matrix and direct integration techniques. Studying proper curvature collineations in each non static case of the above space-times it is shown that when the above space-times admit proper curvature collineations, they form an infinite dimensional vector space.

## 1. INTRODUCTION

This paper is an attampt to investigate the existence of proper curvature collineations (CCS) in the non static plane symmetric space-times. These curvature collineations which preserve the curvature structure of a space-time carry significant information and play an important role in Einstein's theory of general relativity and gravitation. The theory of general relativity, which is actually a field theory of gravitaion and is described in terms of geometry, is highly non-linear [1]. Due to this non-linearaity it becomes very hard to solve the gravitational field equations unless certain symmetry restrictions are imposed on the space-times. These symmetry restrictions may be expressed in terms of Killing vector fields (KVF), homothetic vector fields (HVF), Ricci collineations (RCS) and curvature collineations. Killing vector fields give rise to some conservation laws. Katzin et al. [2,3] suggests that Riemann curvature tensor may also provide some extra understandings which are not provided by (KVF) and (HVF). It is, therefore, important to study CCS. Here an aproach, which is given in [4], is adopted to study proper curvature collineations in non-static plane symmetric space-times by using the rank of the $6\times 6$ Rieman matrix and direct integration techinques. In this paper we are only interested in the non static cases of the above space-times. The cases when the above space-times become static their CCS can be found in [5].

Throughout $M$ represents a four dimensional, connected, Hausdorff space-time manifold with Lorentz metric $g$ of signature (-, +, +, +). The curvature tensor associated with $g_{ab}$, through the Levi-Civita connection, is denoted in component form by $R^a{}_{bcd}$, and the Ricci tensor components are $R_{ab} = R^c{}_{acb}$. The usual covariant, partial and Lie derivatives are denoted by a semicolon, a comma and the symbol $L$, respectively. Round and square brackets denote the usual symmetrization and skew-symmetrization, respectively. Here, $M$ is assumed non-flat in the sense that the curvature tensor does not vanish over any non-empty open subset of $M$.

The covariant derivative of any vector field $X$ on $M$ can be decomposed as

$$X_{a;b} = \frac{1}{2}h_{ab} + F_{ab} \tag{1}$$

where $h_{ab}(=h_{ba}) = L_X g_{ab}$ is a symmetric and $F_{ab}(=-F_{ba})$ is a skew symmetric tensor on $M$. If $h_{ab;c} = 0$, $X$ is said to be *affine* and further satisfies $h_{ab} = 2cg_{ab}, c \in R$ then $X$ is said to be *homothetic* (and *Killing* if $c = 0$). The vector field $X$ is said to be proper affine if it is not homothetic vector field and also $X$ is said to be proper homothetic vector field if it is not Killing vector field.

A vector field $X$ on $M$ is called a curvature collineation (CC) if it satisfies [2]

$$L_X R^a{}_{bcd} = 0 \tag{2}$$

or equivalently,

$$R^a{}_{bcd;e} X^e + R^a{}_{ecd} X^e{}_{;b} + R^a{}_{bed} X^e{}_{;c} + R^a{}_{bce} X^e{}_{;d} - R^e{}_{bcd} X^a{}_{;e} = 0.$$

The vector field $X$ is said to be proper CC if it is not affine [4] on $M$. One can expand the above equation in a set of 22 coupled CC equations which can be seen in [6].

**2. CLASSIFICATION OF THE RIEMANN TENSORS**

In this section we will classify the Riemann tensors in terms of its rank and bivector decomposition.

The rank of the Riemann tensor is the rank of the $6 \times 6$ symmetric matrix derived in a well known way [4]. The rank of the Riemann tensor at $p$ is the rank of the linear map $\beta$ which maps the vector space of all bivectors $F$ at $p$ to itself and is defined by $\beta : F^{ab} \to R^{ab}{}_{cd} F^{cd}$. Define also the subspace $N_p$ of the tangent space $T_p M$ consisting of those members $k$ of $T_p M$ which satisfy the relation

$$R_{abcd} k^d = 0 \qquad (3)$$

Then the Riemann tensor at $p$ satisfies exactly one of the following algebraic conditions [4].

**Class B**

The rank is 2 and the range of $\beta$ is spanned by the dual pair of non-null simple bivectors and $\dim N_p = 0$. The Riemann tensor at $p$ takes the form

$$R_{abcd} = \alpha\, F_{ab} F_{cd} + \eta\, \overset{*}{F}_{ab} \overset{*}{F}_{cd} \qquad (4)$$

where $F$ and its dual $\overset{*}{F}$ are the (unique up to scaling) simple non-null spacelike and timelike bivectors in the range of $\beta$, respectively and $\alpha, \eta \in R$.

**Class C**

The rank is 2 or 3 and there exists a unique (up to scaling) solution say, $k$ of (3) (and so $\dim N_p = 1$). The Riemann tensor at $p$ takes the form

$$R_{abcd} = \sum_{i,j=1}^{3} \alpha_{ij} F^i{}_{ab} F^j{}_{cd} \qquad (5)$$

where $\alpha_{ij} \in R$ for all $i, j$ and $F^i{}_{ab} k^b = 0$ for each of the bivectors $F^i$ which span the range of $\beta$.

**Class D**

Here the rank of the curvature matrix is 1. The range of the map $\beta$ is spanned by a single bivector $F$, say, which has to be simple because the symmetry of Riemann tensor $R_{a[bcd]} = 0$ means $F_{a[b} F_{cd]} = 0$. Then it follows from a standard result that $F$ is simple. The curvature tensor admits exactly two independent solutions $k, u$ of equation (3) so that $\dim N_p = 2$. The Riemann tensor at $p$ takes the form

$$R_{abcd} = \alpha F_{ab} F_{cd} \tag{6}$$

where $\alpha \in R$ and $F$ is simple bivector with blade orthogonal to $k$ and $u$.

**Class O**

The rank of the curvature matrix is 0 (so that $R_{abcd} = 0$) and dim $N_p = 4$.

**Class A**

The Riemann tensor is said to be of class A at $p$ if it is not of class B, C, D or O. Here always dim $N_p = 0$.

A study of the CCS for the above classes A, B, C, D, O and CCS in the two dimensional submanifolds can be found in [4].

## 3. MAIN RESULTS

Consider a non static plane symmetric space-time in the usual coordinate system $(t, x, y, z)$ (labeled by $(x^0, x^1, x^2, x^3)$, respectively) with line element [7]

$$ds^2 = -e^{A(t,x)} dt^2 + e^{B(t,x)} dx^2 + e^{C(t,x)} (dy^2 + dz^2). \tag{7}$$

The Ricci tensor Segre type of the above space-time is {1,1(11)} or {2(11)} or one of its degeneracies. The above space-time admits three linearly independent Killing vector fields which are

$$\frac{\partial}{\partial y}, \frac{\partial}{\partial z}, y\frac{\partial}{\partial z} - z\frac{\partial}{\partial y}. \tag{8}$$

The non-zero independent components of the Riemann tensor are

$$R_{0101} = \frac{1}{4} \left[ \begin{array}{l} e^{A(t,x)} (A_x^2(t,x) + 2 A_{xx}(t,x) - A_x(t,x) B_x(t,x)) - e^{B(t,x)} \\ (B_t^2(t,x) + 2 B_{tt}(t,x) - A_t(t,x) B_t(t,x)) \end{array} \right] \equiv \alpha_1,$$

$$R_{0202} = R_{0303} = -\frac{1}{4} e^{C(t,x)-B(t,x)} \left[ \begin{array}{l} e^{B(t,x)} (C_t^2(t,x) + 2 C_{tt}(t,x) - A_t(t,x) C_t(t,x)) - \\ e^{A(t,x)} A_x(t,x) C_x(t,x)) \end{array} \right] \equiv \alpha_2,$$

$$R_{1212} = R_{1313} = -\frac{1}{4} e^{C(t,x)-A(t,x)} \left[ \begin{array}{l} e^{A(t,x)} (C_x^2(t,x) + 2 C_{xx}(t,x) - B_x(t,x) C_x(t,x)) - \\ e^{B(t,x)} B_t(t,x) C_t(t,x) \end{array} \right] \equiv \alpha_3,$$

$$R_{2323} = -\frac{1}{4} e^{A(t,x)+B(t,x)+2C(t,x)} \left[ e^{A(t,x)} C_x^2(t,x) - e^{B(t,x)} C_t^2(t,x) \right] \equiv \alpha_4,$$

$$R_{0212} = R_{0313} = \frac{1}{4} e^{C(t,x)} \left[ \begin{array}{l} (C_t(t,x) C_x(t,x) + 2 C_{tx}(t,x) - A_x(t,x) C_t(t,x) - \\ B_t(t,x) C_x(t,x)) \end{array} \right] \equiv \alpha_5.$$

Writing the curvature tensor with components $R_{abcd}$ at $p$ as a $6 \times 6$ symmetric matrix

$$R_{abcd} = \begin{pmatrix} \alpha_1 & 0 & 0 & 0 & 0 & 0 \\ 0 & \alpha_2 & 0 & \alpha_5 & 0 & 0 \\ 0 & 0 & \alpha_2 & 0 & \alpha_5 & 0 \\ 0 & \alpha_5 & 0 & \alpha_3 & 0 & 0 \\ 0 & 0 & \alpha_5 & 0 & \alpha_3 & 0 \\ 0 & 0 & 0 & 0 & 0 & \alpha_4 \end{pmatrix}. \quad (9)$$

It is important to remind the reader that we will consider Riemann tensor components as $R^a{}_{bcd}$ for calculating CCS. Here, we are only interested in those cases when the rank of the $6 \times 6$ Riemann matrix is less than or equal to three. Since we know from theorem [4] that when the rank of the $6 \times 6$ Riemann matrix is greater than three there exists no proper CCS. It is also important to remind the reader again that we will only considering the non static cases. The cases when the space-times become static their CCS can be found in [5]. Thus there exist the following possibilities:

(1) Rank=3  $A_x(t,x) = 0$, $B_t(t,x) = 0$, $C_x(t,x) = 0$, $A_t(t,x) \neq 0$, $B_x(t,x) \neq 0$, $C_t(t,x) \neq 0$ and $C_t^2(t,x) + 2C_{tt}(t,x) \neq 0$.

(2) Rank=3  $A_x(t,x) = 0$, $B_t(t,x) = 0$, $B_x(t,x) = 0$, $C_x(t,x) = 0$, $A_t(t,x) \neq 0$, $C_t(t,x) \neq 0$ and $C_t^2(t,x) + 2C_{tt}(t,x) \neq 0$.

(3) Rank=3,  $A_x(t,x) = 0$, $B_t(t,x) = 0$, $C_x(t,x) = 0$, $A_t(t,x) \neq 0$, $B_x(t,x) \neq 0$, $C_t(t,x) \neq 0$ and $C_t^2(t,x) + 2C_{tt}(t,x) = 0$.

(4) Rank=3,  $A_x(t,x) = 0$, $A_t(t,x) = 0$, $B_t(t,x) = 0$, $C_x(t,x) = 0$, $B_x(t,x) \neq 0$, $C_t(t,x) \neq 0$ and $C_t^2(t,x) + 2C_{tt}(t,x) \neq 0$.

(5) Rank=3,  $A_x(t,x) = 0$, $A_t(t,x) = 0$, $B_t(t,x) = 0$, $B_x(t,x) = 0$, $C_x(t,x) = 0$, $C_t(t,x) \neq 0$ and $C_t^2(t,x) + 2C_{tt}(t,x) \neq 0$.

(6) Rank=3,  $A_x(t,x) = 0$, $A_t(t,x) = 0$, $C_x(t,x) = 0$, $B_t(t,x) \neq 0$, $B_x(t,x) \neq 0$, $C_t(t,x) \neq 0$ and $C_t^2(t,x) + 2C_{tt}(t,x) = 0$, $B_t^2(t,x) + 2B_{tt}(t,x) = 0$.

(7) Rank=3,  $A_x(t,x) = 0$, $A_t(t,x) = 0$, $B_x(t,x) = 0$, $C_x(t,x) = 0$, $B_t(t,x) \neq 0$, $C_t(t,x) \neq 0$, $C_t^2(t,x) + 2C_{tt}(t,x) = 0$ and $B_t^2(t,x) + 2B_{tt}(t,x) = 0$.

(8) Rank=3,  $A_x(t,x) = 0$, $A_t(t,x) = 0$, $B_t(t,x) = 0$, $B_x(t,x) \neq 0$, $C_x(t,x) \neq 0$, $C_t(t,x) \neq 0$, $C_t^2(t,x) + 2C_{tt}(t,x) = 0$, $C_t(t,x)C_x(t,x) + 2C_{tx}(t,x) = 0$ and $C_x^2(t,x) + 2C_{xx}(t,x) \neq 0$.

(9) Rank=3, $A_x(t,x) = 0$, $A_t(t,x) = 0$, $B_t(t,x) = 0$, $B_x(t,x) \neq 0$, $C_x(t,x) \neq 0$, $C_t(t,x) \neq 0$, $C_t^2(t,x) + 2C_{tt}(t,x) = 0$, $C_t(t,x)C_x(t,x) + 2C_{tx}(t,x) = 0$ and $C_x^2(t,x) + 2C_{xx}(t,x) = 0$.

(10) Rank=3, $A_x(t,x) = 0$, $A_t(t,x) = 0$, $B_t(t,x) = 0$, $B_x(t,x) = 0$, $C_x(t,x) \neq 0$, $C_t(t,x) \neq 0$, $C_t^2(t,x) + 2C_{tt}(t,x) = 0$, $C_t(t,x)C_x(t,x) + 2C_{tx}(t,x) = 0$ and $C_x^2(t,x) + 2C_{xx}(t,x) \neq 0$.

(11) Rank=3, $A_x(t,x) = 0$, $B_t(t,x) = 0$, $B_x(t,x) = 0$, $A_t(t,x) \neq 0$, $C_x(t,x) \neq 0$, $C_t(t,x) \neq 0$, $C_x^2(t,x) + 2C_{xx}(t,x) = 0$, $C_t(t,x)C_x(t,x) + 2C_{tx}(t,x) = 0$ and $C_t^2(t,x) + 2C_{tt}(t,x) = 0$.

(12) Rank=3, $A_x(t,x) = 0$, $B_t(t,x) = 0$, $B_x(t,x) = 0$, $A_t(t,x) \neq 0$, $C_x(t,x) \neq 0$, $C_t(t,x) \neq 0$, $C_t^2(t,x) + 2C_{tt}(t,x) \neq 0$, $C_t(t,x)C_x(t,x) + 2C_{tx}(t,x) = 0$ and $C_x^2(t,x) + 2C_{xx}(t,x) = 0$.

(13) Rank=3, $A_x(t,x) = 0$, $A_t(t,x) = 0$, $B_x(t,x) = 0$, $B_t(t,x) = 0$, $C_x(t,x) \neq 0$, $C_t(t,x) \neq 0$, $C_t^2(t,x) + 2C_{tt}(t,x) \neq 0$, $C_t(t,x)C_x(t,x) + 2C_{tx}(t,x) = 0$ and $C_x^2(t,x) + 2C_{xx}(t,x) = 0$.

(14) Rank=1, $C_t(t,x) = 0$, $C_x(t,x) = 0$, $A_x(t,x) \neq 0$, $A_t(t,x) \neq 0$, $B_t(t,x) \neq 0$, $B_x(t,x) \neq 0$, $A_x^2(t,x) + 2A_{xx}(t,x) - A_x(t,x)B_x(t,x) \neq 0$ and $B_t^2(t,x) + 2B_{tt}(t,x) - A_t(t,x)B_t(t,x) \neq 0$.

(15) Rank=1, $A_x(t,x) = 0$, $C_t(t,x) = 0$, $C_x(t,x) = 0$, $A_t(t,x) \neq 0$, $B_t(t,x) \neq 0$, $B_x(t,x) \neq 0$ and $B_t^2(t,x) + 2B_{tt}(t,x) - A_t(t,x)B_t(t,x) \neq 0$.

(16) Rank=1, $A_x(t,x) = 0$, $B_x(t,x) = 0$, $C_t(t,x) = 0$, $C_x(t,x) = 0$, $A_t(t,x) \neq 0$, $B_t(t,x) \neq 0$ and $B_t^2(t,x) + 2B_{tt}(t,x) - A_t(t,x)B_t(t,x) \neq 0$.

(17) Rank=1, $C_t(t,x) = 0$, $C_x(t,x) = 0$, $A_x(t,x) \neq 0$, $A_t(t,x) \neq 0$, $B_x(t,x) \neq 0$, $B_t(t,x) \neq 0$, $B_t^2(t,x) + 2B_{tt}(t,x) - A_t(t,x)B_t(t,x) \neq 0$ and $A_x^2(t,x) + 2A_{xx}(t,x) - A_x(t,x)B_x(t,x) = 0$.

(18) Rank=1, $C_t(t,x) = 0$, $C_x(t,x) = 0$, $A_x(t,x) \neq 0$, $A_t(t,x) \neq 0$, $B_t(t,x) \neq 0$, $B_x(t,x) \neq 0$, $A_x^2(t,x) + 2A_{xx}(t,x) - A_x(t,x)B_x(t,x) \neq 0$ and $B_t^2(t,x) + 2B_{tt}(t,x) - A_t(t,x)B_t(t,x) = 0$.

(19) Rank=1, $C_t(t,x) = 0$, $C_x(t,x) = 0$, $A_x(t,x) \neq 0$, $A_t(t,x) \neq 0$, $B_x(t,x) \neq 0$, $B_t(t,x) \neq 0$, $A_x^2(t,x) + 2A_{xx}(t,x) = 0$ and $B_t^2(t,x) + 2B_{tt}(t,x) - A_t(t,x)B_t(t,x) \neq 0$.

(20) Rank=1, $C_t(t,x) = 0$, $C_x(t,x) = 0$, $A_x(t,x) \neq 0$, $A_t(t,x) \neq 0$, $B_x(t,x) \neq 0$, $B_t(t,x) \neq 0$, $A_x^2(t,x) + 2A_{xx}(t,x) - A_x(t,x)B_x(t,x) \neq 0$ and $B_t^2(t,x) + 2B_{tt}(t,x) = 0$.

(21) Rank=1, $A_t(t,x) = 0$, $B_x(t,x) = 0$, $C_x(t,x) = 0$, $C_t(t,x) = 0$, $A_x(t,x) \neq 0$, $B_t(t,x) \neq 0$, $A_x^2(t,x) + 2A_{xx}(t,x) \neq 0$ and $B_t^2(t,x) + 2B_{tt}(t,x) \neq 0$.

(22) Rank=1, $A_t(t,x) = 0$, $B_x(t,x) = 0$, $C_x(t,x) = 0$, $C_t(t,x) = 0$, $A_x(t,x) \neq 0$, $B_t(t,x) \neq 0$ and $A_x^2(t,x) + 2A_{xx}(t,x) = 0$. $B_t^2(t,x) + 2B_{tt}(t,x) \neq 0$.

(23) Rank=1, $A_x(t,x) = 0$, $A_t(t,x) = 0$, $B_x(t,x) = 0$, $C_t(t,x) = 0$, $C_x(t,x) = 0$, $B_t(t,x) \neq 0$ and $B_t^2(t,x) + 2B_{tt}(t,x) \neq 0$.

(24) Rank=1, $A_x(t,x) = 0$, $B_x(t,x) = 0$, $C_t(t,x) = 0$, $C_x(t,x) = 0$, $A_t(t,x) \neq 0$, $B_t(t,x) \neq 0$ and $B_t^2(t,x) + 2B_{tt}(t,x) = 0$.

(25) Rank=1 $A_x(t,x) = 0$, $B_x(t,x) \neq 0$, $C_t(t,x) = 0$, $C_x(t,x) = 0$, $A_t(t,x) \neq 0$, $B_t(t,x) \neq 0$ and $B_t^2(t,x) + 2B_{tt}(t,x) = 0$.

(26)    Rank=1,    $C_t(t,x) = 0$, $C_x(t,x) = 0$, $A_x(t,x) \neq 0$, $A_t(t,x) \neq 0$, $B_x(t,x) \neq 0$, $B_t(t,x) \neq 0$,
$A_x^2(t,x) + 2A_{xx}(t,x) = 0$ and $B_t^2(t,x) + 2B_{tt}(t,x) = 0$.

(27) Rank=1, $A_x(t,x) = 0$, $A_t(t,x) = 0$, $B_t(t,x) = 0$, $B_x(t,x) = 0$, $C_x(t,x) = 0$, $C_t(t,x) \neq 0$ and $C_t^2(t,x) + 2C_{tt}(t,x) = 0$.

(28)    Rank=1,    $A_x(t,x) = 0$, $A_t(t,x) = 0$, $B_t(t,x) = 0$, $B_x(t,x) = 0$, $C_t(t,x) \neq 0$, $C_x(t,x) \neq 0$,
$C_x^2(t,x) + 2C_{xx}(t,x) = 0$, $C_t^2(t,x) + 2C_{tt}(t,x) = 0$ and $C_t C_x(t,x) + 2C_{tx}(t,x) = 0$.

We will consider each case in turn.

**Case 1**

In this case we have $A_x(t,x) = 0$, $B_t(t,x) = 0$, $C_x(t,x) = 0$ $A_t(t,x) \neq 0$, $B_x(t,x) \neq 0$, $C_t(t,x) \neq 0$, $C_t^2(t,x) + 2C_{tt}(t,x) \neq 0$ and the rank of the $6 \times 6$ Riemann matrix is three. Here, there exists a unique (up to a multiple) no where zero spacelike vector field $x_a = x_{,a}$ satisfying $x_{a;b} = 0$. From the Ricci identity $R^a{}_{bcd} x_a = 0$. From the above constraints we have $A_x(t,x) = 0$, $B_t(t,x) = 0$ and $C_x(t,x) = 0 \Rightarrow A(t,x) = \alpha(t)$, $B(t,x) = \beta(x)$ and $C(t,x) = \eta(t)$. Substituting the information of $A(t,x)$, $B(t,x)$ and $C(t,x)$ in (7) and after a rescaling of $x$, the line element can be written in the form

$$ds^2 = -e^{\alpha(t)}dt^2 + dx^2 + e^{\eta(t)}(dy^2 + dz^2). \tag{10}$$

The above space-time (10) is 1+3 decomposable and belongs to curavture class C. CCS in this case [4] are

$$X = f(x)\frac{\partial}{\partial x} + X', \tag{11}$$

where $f(x)$ is an arbitrary function of $x$ and $X'$ is a homothetic vector field in the induced geometry on each of the three dimensional submanifolds of constant $x$. The completion of case 1 requires finding the homothetic vector fields in the induced geometry on the submanifolds of constant $x$. The induced metric $g_{\lambda\omega}$ (where $\lambda, \omega = 0,2,3$) has non zero components given by

$$g_{00} = -e^{\alpha(t)}, \qquad g_{22} = g_{33} = e^{\eta(t)} \tag{12}$$

A vector field $X'$ is called a homothetic vector field if it satisfies

$$L_{X'} g_{\alpha\beta} = c g_{\alpha\beta}, \quad c \in R. \tag{13}$$

One can expand (13) using (12) to get

$$\dot{\alpha} X^0 + 2X^0{}_{,0} = c, \tag{14}$$

$$-e^{\alpha} X^{0}{}_{,2} + e^{\eta} X^{2}{}_{,0} = 0, \tag{15}$$

$$-e^{\alpha} X^{0}{}_{,3} + e^{\eta} X^{3}{}_{,0} = 0, \tag{16}$$

$$\dot{\eta} X^{0} + 2 X^{2}{}_{,2} = c, \tag{17}$$

$$X^{2}{}_{,3} + X^{3}{}_{,2} = 0, \tag{18}$$

$$\dot{\eta} X^{0} + 2 X^{3}{}_{,3} = c. \tag{19}$$

Equations (14), (15) and (16) give

$$X^{0} = \frac{1}{2} c e^{-\frac{\alpha}{2}} \int e^{\frac{\alpha}{2}} dt + e^{-\frac{\alpha}{2}} N^{1}(y,z), \quad X^{2} = N^{1}_{y}(y,z) \int e^{\frac{\alpha}{2}-\eta} dt + N^{2}(y,z),$$

$$X^{3} = N^{1}_{z}(y,z) \int e^{\frac{\alpha}{2}-\eta} dt + N^{3}(y,z),$$

where $N^{1}(y,z)$, $N^{2}(y,z)$, and $N^{3}(y,z)$ are functions of integrartion. If one proceeds further, after a straightforward calculation one can find that the proper homothetic vector fields exist if and only if $\eta(t) = \ln(at+b)^{2}$ and $\alpha(t) = d$, where $a, b, d \in R$. Substituting this information into (9), one finds that the rank of the Riemann matrix reduces to one, thus giving a contradiction. So the only homothetic vector fields in the induced geometry are the Killing vector fields which are

$$X^{0} = 0, \quad X^{2} = z c_{1} + c_{2}, \quad X^{3} = -y c_{1} + c_{3}, \tag{20}$$

where $c_{1}, c_{2}, c_{3} \in R$. CCS in this case are given by use of equation (11) and (20)

$$X^{0} = 0, \quad X^{1} = f(x), \quad X^{2} = z c_{1} + c_{2}, \quad X^{3} = -y c_{1} + c_{3}, \tag{21}$$

where $f(x)$ is an arbitrary function of $x$. One can write the above equation (21) after subtracting Killing vector fields as

$$X = (0, f(x), 0, 0). \tag{22}$$

CCS clearly form an infinite dimensional vector space. Cases 2 to 5 are precisely the same.

**Case 6**

In this case we have $A_{x}(t,x) = 0$, $A_{t}(t,x) = 0$, $C_{x}(t,x) = 0$, $B_{x}(t,x) \neq 0$, $B_{t}(t,x) \neq 0$, $C_{t}(t,x) \neq 0$  $C_{t}^{2}(t,x) + 2 C_{tt}(t,x) = 0$, $B_{t}^{2}(t,x) + 2 B_{tt}(t,x) = 0$ and rank of the $6 \times 6$ Riemann matrix is three. Here, there exists a unique (up to a multiple) no where zero timelike vector field $t_{a} = t_{,a}$ which is the solution of equation (3). The vector field $t_{a}$ is not covariantly constant. From the above constraints we have $A_x(t,x) = 0$, $A_t(t,x) = 0$ and $C_x(t,x) = 0 \Rightarrow A(t,x) = b$ and $C(t,x) = D(t)$, where $b \in R$ and $D(t)$ is a function of integration. Equations $D_t^2(t) + 2D_{tt}(t) = 0$ and $B_t^2(t,x) + 2B_{tt}(t,x) = 0 \Rightarrow D(t) = \ln(at+d)^2$ and $B(t,x) = \ln(P(x)t + Q(x))$, where $P(x)$ is no where zero functions of integration and $Q(x)$ is function of integration and $a,d \in R (a \neq 0)$. The line element in this case can, after a rescaling of $t$, be written in the form

$$ds^2 = -dt^2 + (P(x)t + Q(x))^2 dx^2 + (at+d)^2(dy^2 + dz^2). \tag{23}$$

The above space-time (23) belongs to curavture class C. Substituting the above information into the CC equations in [6] and after some calculation one finds CCS in this case are

$$X^0 = 0, \quad X^1 = 0, \quad X^2 = zc_1 + c_2, \quad X^3 = -yc_1 + c_3, \tag{24}$$

where $c_1, c_2, c_3 \in R$. CCS in this case are Killing vector fields. Cases 7 to 13 are precisely the same.

**Case 14**

In this case one has $A_t(t,x) \neq 0$, $B_t(t,x) \neq 0$, $C_x(t,x) = 0$, $C_t(t,x) = 0$, $A_x(t,x) \neq 0$, $B_x(t,x) \neq 0$, $A_x^2(t,x) + 2A_{xx}(t,x) - A_x(t,x)B_x(t,x) \neq 0$, $B_t^2(t,x) + 2B_{tt}(t,x) - A_t(t,x)B_t(t,x) \neq 0$ and the rank of the $6 \times 6$ Riemann matrix is one. From the above constraints we have $A = A(t,x)$, $B = B(t,x)$ and $C(t,x) = e$, where $e \in R$. Here there exist two linear independent solutions $y_a = y_{,a}$ and $z_a = z_{,a}$ of equation (3) and satisfying $y_{a;b} = 0$ and $z_{a;b} = 0$. The line element can, after rescaling of $y$ and $z$, be written in the form

$$ds^2 = -e^{A(t,x)}dt^2 + e^{B(t,x)}dx^2 + (dy^2 + dz^2). \tag{25}$$

The above space-time (25) is 1+1+2 decomposable and belongs to curavture class D. CCS in this case are [4]

$$X = f(y,z)\frac{\partial}{\partial y} + g(y,z)\frac{\partial}{\partial z} + X' \tag{26}$$

where $f(y,z)$ and $g(y,z)$ are arbitrary functions of $y$ and $z$ and $X'$ is a CC on each of two dimensional submanifolds of constant $y$ and $z$. The next step is to find the CCS in the induced geometry of the submanifolds of constant $y$ and $z$. A method for finding CCS in 2-dimensional submanifolds is given in [4]. If one proceeds further the non zero components of the induced metric on each of the two dimensional submanifolds of constant $y$ and $z$ are given by

$$g_{00} = -e^{A(t,x)} \quad g_{11} = e^{B(t,x)}. \tag{27}$$

The nonzero components of the Ricci tensor are

$$R_{00} = \frac{1}{4} e^{-B(t,x)} [(A'^2 + 2A'' - A'B') e^{A(t,x)} - e^{B(t,x)} (\dot{B}^2 + 2\ddot{B} - \dot{A}\dot{B})]$$

$$R_{11} = -\frac{1}{4} e^{-A(t,x)} [(A'^2 + 2A'' - A'B') e^{A(t,x)} - e^{B(t,x)} (\dot{B}^2 + 2\ddot{B} - \dot{A}\dot{B})]. \tag{28}$$

Here dot and prime denote the partial derivative with respect to $t$ and $x$, respectively. The Ricci scalar is

$$R = -\frac{1}{2} [(A'^2 + 2A'' - A'B') e^{-B(t,x)} - e^{-A(t,x)} (\dot{B}^2 + 2\ddot{B} - \dot{A}\dot{B})].$$

It follows from [4] that CCS in the two dimensional submanifolds of constant $y$ and $z$ are the solution of the equation $L_{X'} G_{\alpha\beta} = 0$, where $G_{\alpha\beta} \equiv \frac{R}{2} g_{\alpha\beta}$ and $\alpha, \beta = 0, 1$. The non-zero components of $G_{\alpha\beta}$ are

$$G_{00} = \frac{1}{4} e^{-B(t,x)} [(A'^2 + 2A'' - A'B') e^{A(t,x)} - e^{B(t,x)} (\dot{B}^2 + 2\ddot{B} - \dot{A}\dot{B})]$$

$$G_{11} = -\frac{1}{4} e^{-A(t,x)} [(A'^2 + 2A'' - A'B') e^{A(t,x)} - e^{B(t,x)} (\dot{B}^2 + 2\ddot{B} - \dot{A}\dot{B})]. \tag{29}$$

Expanding the equation $L_{X'} G_{\alpha\beta} = 0$ and using equation (29) we get

$$\dot{G}_{00} X^0 + G'_{00} X^1 + 2 G_{00} X^0_{,0} = 0, \tag{30}$$

$$e^{B(t,x)} X^1_{,0} - e^{A(t,x)} X^0_{,1} = 0, \tag{31}$$

$$\dot{G}_{11} X^0 + G'_{11} X^1 + 2 G_{11} X^1_{,1} = 0. \tag{32}$$

Here, the above system of equations give trivial solution which is $X^0 = X^1 = 0$. Proper CCS in this case can be written as

$$X = (0, 0, f(y, z), g(y, z)), \tag{33}$$

Clearly CCS in this case form an infinite dimensional vector space. Cases 15 to 26 are precisely the same.

**Case 27**

In this case we have $A_x(t,x)=0$, $A_t(t,x)=0$, $B_t(t,x)=0$, $B_x(t,x)=0$, $C_x(t,x)=0$, $C_t(t,x)\neq 0$, $C_t^2(t,x)+2C_{tt}(t,x)=0$ and the rank of the $6\times 6$ Riemann matrix is one. From the above constraints we have $A(t,x)=a$, $B(t,x)=b$ and $C(t)=\ln(et+d)^2$, where $a,b,e,d\in R(e\neq 0)$. Here there exist two linear independent solutions $t_a=t_{,a}$ and $x_a=x_{,a}$ of equation (3). The vector field $t_a$ is not a covariantly constant whereas $x_a$ is covariantly constant. Substituting the information of $A(t,x)$, $B(t,x)$ and $C(t,x)$ in (7) and after rescaling of $t$ and $x$, the line element can be written in the form

$$ds^2 = -dt^2 + dx^2 + (et+d)^2(dy^2+dz^2). \tag{34}$$

The above space-time (34) is 1+3 decomposable and belongs to curavture class D. Substituting the above information into CC equations in [6] one finds

$$X^0{}_{,2} = X^0{}_{,3} = 0, \qquad X^1{}_{,2} = X^1{}_{,3} = 0,$$
$$X^2{}_{,0} = X^2{}_{,1} = X^2{}_{,2} = 0, \qquad X^3{}_{,0} = X^3{}_{,1} = X^3{}_{,3} = 0, \tag{35}$$

$$X^2{}_{,3} + X^3{}_{,2} = 0 \tag{36}$$

Equation (35) gives $X^0 = f(t,x)$, $X^1 = g(t,x)$, $X^2 = M(z)$ and $X^3 = N(y)$, where $f(t,x)$ and $g(t,x)$ are arbitrary functions and $M(z)$ and $N(y)$ are functions of integration. Substituting the above information about $X^2$ and $X^3$ in (36) gives $M_z(z)+N_y(y)=0$, upon differentiating with respect to $y$ gives $N_{yy}(y)=0 \Rightarrow N(y)=c_1 y+c_2$, substituting back in the same equation we get $M(z)=-c_1 z+c_3$, where $c_1,c_2,c_3 \in R$. CCS in this case

$$X^0 = f(t,x), \ X^1 = g(t,x), \ X^2 = -c_1 z+c_3, \ X^3 = c_1 y+c_2. \tag{37}$$

One can write the above equation (37) after subtracting Killing vector fields as

$$X = (f(t,x), g(t,x), 0, 0). \tag{38}$$

Clearly CCS form an infinite dimensional vector space.

**Case 28**

In this case we have $A_x(t,x) = 0$, $A_t(t,x) = 0$, $B_t(t,x) = 0$, $B_x(t,x) = 0$, $C_t(t,x) \neq 0$, $C_x(t,x) \neq 0$, $C_x^2(t,x) + 2C_{xx}(t,x) = 0$, $C_x^2(t,x) + 2C_{xx}(t,x) = 0$, $C_t C_x(t,x) + 2C_{tx}(t,x) = 0$ and the rank of the $6 \times 6$ Riemann matrix is one. From the above constraints we have $A(t,x) = a$, $B(t,x) = b$ and $C(t,x) = \ln(ct + ex + d)^2$, where $a, b, e, d \in R (c, e \neq 0)$. Here, there exist two linear independent solutions $t_a = t_{,a}$ and $x_a = x_{,a}$ of equation (3). The vector fields $t_a$ and $x_a$ are not covariantly constant. The line element after rescaling of $t$ and $x$, can be written in the form

$$ds^2 = -dt^2 + dx^2 + (ct + ex + d)^2 (dy^2 + dz^2). \qquad (39)$$

The above space-time (39) belongs to curavture class D. Substituting the above information into the CC equations in [6] and after some calculation one find CCS in this case are Killing vector fields which are given in equation (24).

**SUMMARY**

In this paper a study of non-static plane symmetric space-times according to their proper CCS is given. An approach is adopted to study the above space-times by using the rank of the $6 \times 6$ Riemann matrix and also using the theorem given in [4], which suggested where proper CCS exist. From the above study we obtain the following results:

(i) We get the space-time (10) that admits proper curvature collineations (see case 1) when the rank of the $6 \times 6$ Riemann matrix is three and there exists a nowhere zero independent covariantly constant spacelike vector field, which is the solution of equation (3).

(ii) We obtain the space-time (23) that admits curvature collineations when the rank of the $6 \times 6$ Riemann matrix is three and there exists a unique nowhere zero independent timelike vector field, which is the solution of equation (3) and is not covariantly constant. In this case the curvature collineations are Killing vector fields (for details see case 6).

(iii) The space-time (25) is obtained, which admits proper curvature collineations (see case 14) when the rank of the $6 \times 6$ Riemann matrix is one and there exist two independent covariantly constant spacelike vector fields being the solutions of equation (3).

(iv)    The space-time (34) is obtained, which admits proper curvature collineations (see case 27) when the rank of the $6\times 6$ Riemann matrix is one and there exist two nowhere zero independent solutions of equation (3) of which only one is covariantly constant vector field.

(v)    The space-time (39) is achieved, which admits curvature collineations (see case 28) when the rank of the $6\times 6$ Riemann matrix is one and there exist two nowhere zero independent timelike and spacelike vector fields being the solutions of equation (3) and are not covariantly constant. In this case curvature collineations are Killing vector fields.